\documentstyle[12pt,amsfonts,epsf]{article}

\newcommand{\beq}{\begin{equation}}
\newcommand{\eeq}{\end{equation}}
\newcommand{\ba}{\begin{array}}
\newcommand{\ea}{\end{array}}
\newcommand{\bea}{\begin{eqnarray}}
\newcommand{\eea}{\end{eqnarray}}
\newcommand{\bean}{\begin{eqnarray*}}
\newcommand{\eean}{\end{eqnarray*}}
\newcommand{\nin}{\noindent}
\newcommand{\ra}{\rightarrow}

\newcommand{\xx}{\hbox{\rm x}}

\newcommand{\bx}{{\bold x}}

\newcommand{\bn}{{\bold n}}
\newcommand{\bX}{{\bold X}}

 \newcommand{\RR}{{\Bbb R}}
 \newcommand{\PP}{{\Bbb P}}
\newcommand{\ZZ}{{\Bbb Z}}

\begin{document}
 
\title{
Multidimensional Quadrilateral Lattices\\
are Integrable}

\author{ Adam Doliwa$^{1,2}$\thanks{e-mail: DOLIWA@ROMA1.INFN.IT and 
DOLIWA@FUW.EDU.PL} \/ and Paolo Maria Santini$^{1,3}$\thanks{e-mail: 
SANTINI@CATANIA.INFN.IT and SANTINI@ROMA1.INFN.IT} \\
$^1$ Istituto Nazionale di Fisica Nucleare, Sezione di Roma\\ 
P-le Aldo Moro 2, I--00185 Roma, Italy \\  
$^2$ Instytut Fizyki Teoretycznej, Uniwersytet Warszawski \\
ul. Ho\.{z}a 69, 00-681 Warszawa, Poland \\  
$^3$ Dipartimento di Fisica, Universit\`a di Catania \\
Corso Italia 57, I-95129 Catania, Italy\\  
}

\date{}
\maketitle

\begin{abstract}
\nin The notion of multidimensional quadrilateral lattice is 
introduced. It is shown that such a lattice is characterized by
a system of integrable discrete nonlinear equations. Different useful
formulations of the system are given. The geometric construction of the
lattice is also discussed and, in particular, it is clarified the number
of initial--boundary data which define the lattice uniquely.

\bigskip

\nin {\it Keywords:} Integrable systems, discrete geometry
\end{abstract}


\vskip1cm

Preprint ROME1-1162/96, November 28, 1996

Dipartimento di Fisica, Universit\`a di Roma "La Sapienza"

I.N.F.N. -- Sezione di Roma

\vfill

\pagebreak      

\section{Introduction}
During the last three decades a new branch of mathematical physics, 
the theory
of integrable systems, has been developed \cite{AblSeg}--\cite{Konop}. Besides 
its mathematical beauty, 
the theory has found also an enormous variety of 
applications in physics. 

A lot of attention has been dedicated also to discrete (difference) 
integrable systems 
\cite{AblSeg}\cite{ZMNP}\cite{FadTa}--\cite{KIB}\cite{Esterel}--\cite{CaNij}. 
They share with their differential 
counterparts many common properties, like for example an integrability
scheme, constants of motion, soliton solutions, etc. 

To construct an integrable discretisation of the given soliton equation
one can follow several different approaches. The standard ones 
are: \\
i) a discrete version of the Lax pair (see, for example \cite{AblSeg},
\cite{Esterel}) \\
ii) the Hirota method via a bilinear form \cite{Hir}\cite{DJM} \\
iii) extensions of the Zakharov -- Shabat dressing method 
\cite{LPS}\cite{BoKo} \\
iv) direct linearization using linear integral equations \cite{NQC1}-\cite{CaNij}.

Recently, another discretisation approach has been applied to those 
soliton equations which describe geometrically meaningful objects, like
curves and surfaces. 
It is well known that many integrable PDE's allow for such an interpretation 
(see, for example \cite{Sym}-\cite{Konop1} and references cited there). 
Given a geometric integrable system, the geometry 
provides its integrability 
scheme (the associated linear problem) and
associates with such an 
integrable system other integrable ones, which describe various
objects "living" on the same submanifold \cite{Sym} (see, for instance, the 
connection between the integrable vortex motion and the Heisenberg spin model). 

It may be interesting to note that many
of the important integrable PDE's were studied by the prominent geometers
of the XIXth century \cite{Darb}\cite{Bianchi}. They have found also 
tools which allow to construct explicit formulas for the immersed 
submanifolds, like the Darboux transformation \cite{Darb}.

In the process of discretisation of such systems, as a working principle, 
the discrete 
analogues of the relevant geometric properties must be found. In this way the 
discrete curves 
\cite{DS1} were "incorporated" into the soliton theory, and some classes of 
surfaces, including the pseudospherical, minimal, isothermal and constant 
mean curvature surfaces were discretised \cite{BP1}-\cite{PWu}. Recently 
a new example of integrable discrete surface, which contains in principle
all the ones known
before as a special reductions, has been found and related to the fully
discrete Toda system (Hirota equation) \cite{DCN}.

In this paper we discretise the geometrically meaningful integrable 
multidimensional system which describes submanifolds 
pa\-ra\-me\-tri\-zed by 
multiconjugate coordinates. The associated differential equations were derived  
by Darboux \cite{Darb}
and rediscovered (and solved), in the matrix case, about ten 
years ago by Zakharov and Manakov 
\cite{ZaMa}, using a $\bar{\partial}$ approach; for this reason, they are
now called Darboux--Zakharov--Manakov (DZM) equations.
More precisely, we do the following. \\
i) We show that the proper
geometric discretisation of the notion of (multidimensional) conjugate net
is the notion of (multidimensional) quadrilateral lattice (MQL), meaning
that all the elementary quadrilaterals of the lattice are planar. \\
ii) We characterize MQL's in terms of a system of nonlinear integrable discrete
equations in $N$ dimensions, $N\geq 3$ (equations (\ref{eq:dDarb1}) of 
Section 2). \\
iii) We show that the lattice is defined uniquely by $\frac{N(N-1)}{2}$ 
"initial" surfaces or, equivalently, by $N$ "initial" curves plus $N(N-1)$
arbitrary "initial--boundary" data, which are functions of two variables.

We remark that the equations which characterize the MQL have been
recently obtained by Bogdanov and Konopelchenko \cite{BoKo}
as a natural discrete integrable analogue of the DZM
equations, but their geometric meaning was unknown.

The MQL introduced in this paper allows one to generate, through a systematic
geometrically meaningful reduction procedure, not only all the known geometric
integrable lattices previously mentioned in this introduction, but also new 
ones.
For this reason, a systematic investigation of the reductions of the MQL has
recently started \cite{CDS} and the list of integrable geometric lattices 
already obtained
through it includes the following. \\
i) The fully discrete Toda lattice (Hirota equation) which, as it was shown in 
\cite{DCN}, describes the Laplace sequence of 2-dimensional quadrilateral 
lattices. \\
ii) The discrete analogue of the Lam\'e equations for orthogonal nets, 
corresponding
to the case when the elementary quadrilaters are inscribed in circles. \\
iii) Some symmetric reductions of the MQL including, for example, a discrete
analogue of the $(-1)$ element of the Nizhnik--Veselov--Novikov hierarchy
of $2+1$ dimensional integrable systems.

The integrability of the equations associated with the MQL was established in
\cite{BoKo} using the $\bar{\partial}$ approach. A more geometric way of 
constructing solutions of the MQL equations has been recently obtained
in \cite{DMS}, using a discrete vectorial variant of the Darboux
transformations. These transformations allow to construct, from a given MQL,
a new one, which is, in general, topologically more complicated. In 
particular, dromionic and rational MQL's have been obtained.

The paper is organized as follows. In Section 2 we give the definition of the
multidimensional quadrilateral lattice and derive the underlying integrable 
system of discrete equations. In Section 3 we present  
different forms of the system.
In Section 4 we discuss the initial value problem for the MQL's.

\section{Multidimensional Quadrilateral Lattices}
As it was mentioned in the introduction, the notion of quadrilateral
lattice is the discrete generalization of the notion of conjugate nets.

In the continuous case, conjugate coordinates on a surface are defined
as follows \cite{Eisenhart}. Given a point on a surface and a second point 
which belongs to the 
coordinate line passing through the first one, the tangent planes to the surface
at those two points intersect along one straight line. In the limit when
these two points coincide, the direction of the straight line is the tangent
direction to the second coordinate line (see Fig. 1).

\epsffile{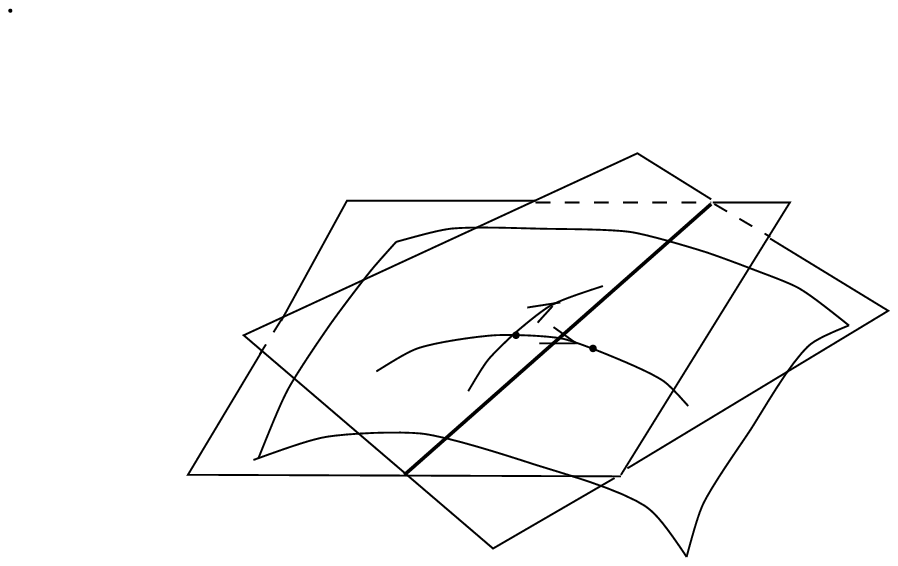}

Figure 1.

\bigskip

On a discrete level, a discrete surface can be defined as a mapping
from $\ZZ^2$ to the $M$ dimensional space $\RR^M$
\beq x:\ZZ^2 \ra \RR^M \; \; .
\eeq 
If we consider the two neighbouring "tangent planes" along the first
coordinate line, defined by the triples $\langle x,T_1 x, T_2 x \rangle, \;
\langle T_1 x,T_1^2 x, T_1 T_2 x \rangle$, where  $T_i$ is the shift operator 
in the $i$-th direction of the lattice, their intersection line 
coincides with the direction of the second coordinate line if and only if 
the elementary quadrilateral $\{x,T_1 x, T_2 x, T_1 T_2 x \}$ is  planar
(see Fig. 2).

\epsffile{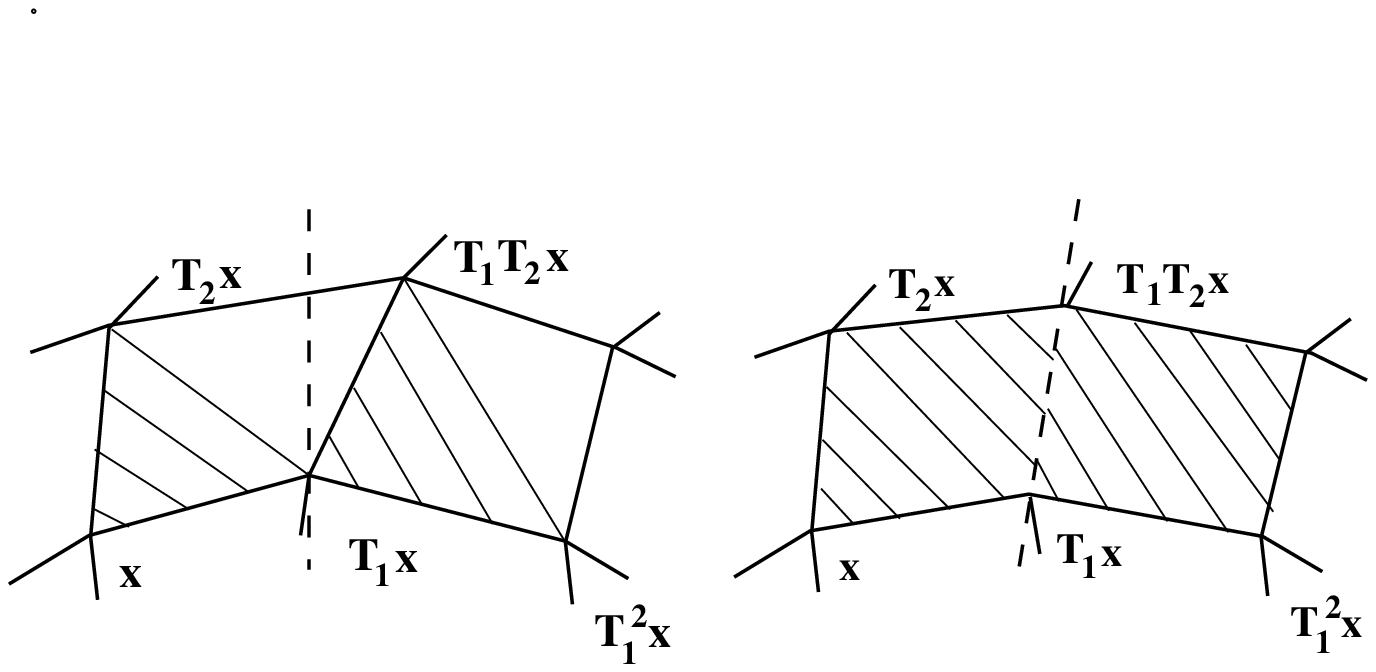}

Figure 2.

\bigskip

\nin Therefore the natural discrete analogue of the notion of 2-dimensional
conjugate net is given by the notion of 2-dimensional quadrilateral lattice 
\cite{Sauer}.

\bigskip

\nin {\bf Definition 1} A 2-dimensional quadrilateral lattice is a mapping
$x:\ZZ^2 \ra \RR^M \; , \; M\geq 2 $ such that all the elementary quadrilaterals
$\{x,T_1 x, T_2 x, T_1 T_2 x \}$ are planar.

\bigskip

A multidimensional conjugate net is characterized by the property that any 
surface made by varying two parameters forms a 2-dimensional conjugate net 
with respect
to these two coordinates. The discrete analogue of the above notion is therefore
given by the notion of multidimensional quadrilateral lattice.
\bigskip

\nin {\bf Definition 2} By an $N$-dimensional quadrilateral lattice we mean 
a mapping from an 
$N$-dimensional integer lattice to the $M$-dimensional linear space
\beq
 x : \ZZ^N\ra \RR^M \; \; , \, M\geq N 
\eeq
such that all the elementary quadrilaterals with vertices 
\beq \{x,T_i x, T_j x, T_i T_j x \} \; \; \; , \;  \; i,j = 1 \dots N \; \; , 
\; \; i\not= j \; \; ,
\eeq
are planar.

Since the planarity of four points is preserved during the projective
transformations of the ambient space, quadrilateral lattices are objects 
which actually should be described whithin the projective geometry approach. 
In the above definition, therefore, it is very convenient to replace the 
linear space $\RR^M$ by the projective space $\PP^M$:
\beq
 x : \ZZ^N\ra \PP^M \; \; , \, M\geq N \; \; . 
\eeq
To perform the calculations it is natural then to use homogeneous
coordinates $\bx\in\RR^{M+1}\setminus\{0\}$ such that $x=[\bx]\in\PP^M$
are the corresponding directions.

In terms of the homogeneous coordinates the {\it linear} relations between 
the vectors representing
vertices of the elementary quadrilaterals can be written as a system of 
$\frac{N(N-1)}{2}$ discrete Laplace equations
\beq \label{eq:dpLap}
D_iD_j\bx = (T_i A_{ij})D_i\bx + (T_j A_{ji})D_j\bx + C_{ij}\bx
\; \; , \; \; i\not= j , \; \; i,j=1,\dots , N \; \; , \; \; 
\eeq
where $D_i=T_i - 1$ is the discrete partial derivative, $A_{ij}$ and 
$C_{ij}=C_{ji}$ are scalar
functions of the lattice variable 
$\bn = (n_1,\dots , n_N)\in \ZZ^N$:
\beq A_{ij} \; , \; C_{ij}: \ZZ^N \ra \RR \;
\; \; , \; \; i\not= j , \; \; i,j=1,\dots , N \; \;   .
\eeq
The projective picture does not change if we multiply the homogeneous
coordinates by a scalar function
\beq \bx \ra \tilde{\bx} = \frac{1}{\rho}\bx \; \; \; \; , \; \; \; \; \; \;
\rho :\ZZ^N \ra \RR\setminus \{ 0 \} \; \;  .
\eeq
In fact, after such a gauge transformation, the new coordinates satisfy the 
Laplace equation (\ref{eq:dpLap}) with the coefficients
\bea \tilde{A}_{ij} & = & \frac{1}{T_j\rho}(A_{ij}\rho - D_j\rho ) 
\; \; \; \; \; , \; \;  \; \; i\not= j , \; \; i,j=1,\dots , N \; \; , \; \; \\
\tilde{C}_{ij} & = & \frac{1}{T_iT_j\rho}(-D_iD_j\rho + (T_iA_{ij})D_i\rho +
(T_jA_{ji})D_j\rho + C_{ij}\rho ) \; \; . \nonumber
\eea
Taking as gauge function (say) the last coordinate of $\bx$, i.e. 
$\rho =\xx^{M+1}$ (the non-homogeneous gauge), we "place" our net in the 
$M$-dimensional affine subspace of $\RR^{M+1}$ characterized by
$\xx^{M+1}=1$. Then we get the special Laplace equations 
\beq \label{eq:dLap}
D_iD_j\bx = (T_i A_{ij})D_i\bx + (T_j A_{ji})D_j\bx \; \; , \; \; i\not= j , 
\; \; i,j=1,\dots , N \; ,
\eeq
where we can identify also the affine subspace with $\RR^M$, i.e. in the
above formula, $\bx=(\xx^1,...,\xx^M)\in\RR^M$.

For $N=2$ we have only one linear equation and no restrictions on its 
coefficients. The solution  of the
Laplace equation represents then a two-dimensional quadrilateral lattice (or 
discrete conjugate net) \cite{DCN}.

Starting from $N=3$, we must repeate the two-dimensional construction 
in every pair of directions and, to make it possible, we should take
into account the following compatibility condition between the Laplace 
equations 
(\ref{eq:dpLap})
\bea \label{eq:dpDarb1}
D_kA_{ij} & = & A_{ij}T_j A_{jk} +
A_{ik}T_k A_{kj} - A_{ik}T_k A_{ij} + C_{ij} \; \; , \\
D_k C_{ij} & = & D_i C_{kj} \; \; \; \; \; \; \; \; \; \;, \; \; \; \; \; 
i\not=j\not= k \not= i\;  .   \nonumber
\eea
In the non-homogeneous gauge $(C_{ij}=0)$ the above equations reduce to 
$N(N-1)(N-2)$ equations for the $N(N-1)$ coefficients $A_{ij}$
\beq \label{eq:dDarb1}
D_kA_{ij} = A_{ij}T_j A_{jk} + A_{ik}T_k A_{kj} -  A_{ik}T_k A_{ij} \; ,
\; \; i\not=j\not= k  \not= i\; .
\eeq
Therefore the system of equations (\ref{eq:dDarb1}) is overdetermined for $N>3$
and determined only for $N=3$.

The continuous limit of the equations (\ref{eq:dLap}) and (\ref{eq:dDarb1}) 
gives rise to the classical Laplace equations 
\beq \label{eq:cLap}
\bx_{,ij} = a_{ij}\bx_{,i} + a_{ji} \bx_{,j} \; \; , \; \; i\not=j \; \; ,
\eeq
and to the DZM equations for conjugate multidimensional nets \cite{Darb}
\beq \label{eq:cDarb1}
a_{ij,k} = a_{ij}a_{jk} + a_{ik}a_{kj} - a_{ik}a_{ij} \; \; ,
\; \; \; i\not=j\not= k  \not= i\; ,
\eeq
where $f_{,i}=\frac{\partial f}{\partial u_i}$.

We conclude this Section by pointing out that multidimensional quadrilateral 
lattices are 
characterized by the set of integrable equations (\ref{eq:dDarb1}), and can 
be therefore called {\it integrable lattices}. Their integrability scheme
is governed by the {\it linear} (discrete) Laplace equations (\ref{eq:dpLap})
(or (\ref{eq:dLap})) which express the planarity of the elementary 
quadrilaterals.

In the rest of the paper we will refer to equations (\ref{eq:dDarb1}) (or 
(\ref{eq:dpDarb1})) as to the multidimensional quadrilateral lattice 
equations.

\section{Alternative Formulations of the Multidimensional Quadrilateral Lattice
Equations}
In this Section we discuss, for completness, other forms of the MQL equations 
which will be usefull in deriving integrable reductions \cite{CDS}
and in the construction of the Darboux transformations \cite{DMS} for the lattice.

Equations (\ref{eq:dpDarb1}b), together with the symmetry condition 
$C_{ij}=C_{ji}$, imply the 
existence of a function $C:\ZZ^N \ra \RR$ such that
\beq C_{ij} = D_iD_j C \; \; \; \; i\not= j .
\eeq
On the other hand from equations (\ref{eq:dpDarb1}a) one can derive the 
following constraint
\beq 
\frac{T_k(A_{ji}+1)}{A_{ji}+1} = \frac{T_i(A_{jk}+1)}{A_{jk}+1} \; \; ,
 \; \; \; i\not= j \not= k \not= i 
\eeq  
which is automatically satisfied expressing $A_{ij}$ in terms of the 
"logarithmic potentials" 
$\gamma_i:\ZZ^N\ra\RR$ 
\beq \label{def:gamma}
A_{ij} = \frac{D_j\gamma_i}{\gamma_i} \; \; \; , \; \; \;  i \not= j \; \; .
\eeq
Finally, the nonlinear system (\ref{eq:dpDarb1}) can be rewritten as a system
of $\frac{N(N-1)(N-2)}{2}$ equations
\beq \label{eq:dpDarb2}
D_kD_j\gamma_i = \left( T_j\frac{D_k\gamma_j}{\gamma_j}\right)D_j\gamma_i +
\left( T_k\frac{D_j\gamma_k}{\gamma_k}\right)D_k\gamma_i + \gamma_i D_iD_jC \; 
\; \; , \; i\not=j\not= k \not=i \; .
\eeq
We remark that the last term vanishes when working in the non-homogeneous
coordinates.

To see the geometric meaning of the function $\gamma_i$ defined in 
(\ref{def:gamma}) let us note 
that, when taken as a gauge function, it removes all the corresponding
coefficients $A_{ij} \; , (j=1,\dots , N \; , \; j\not= i)$ from the Laplace 
equations (\ref{eq:dpLap}).

Of course, the definition of the functions $\gamma_i$ depends on the 
particular gauge we start with.
>From now on, we fix the non-homogeneous gauge (\ref{eq:dLap}) and define
the functions $\gamma_i$ with respect to that gauge. However, we still 
have the freedom 
of multiplying $\gamma_i$ by an arbitrary function of the single variable
$n_i$.  
                                                             
\bigskip

\nin Let us rewrite the special Laplace equations (\ref{eq:dLap}) using the 
functions
$\gamma_i$
\beq \label{eq:dLap2}
D_iD_j\bx= \left( T_i\frac{D_j\gamma_i}{\gamma_i}\right)D_i\bx +
\left( T_j\frac{D_i\gamma_j}{\gamma_j}\right)D_j\bx \; \; , \; \; i\not=j \; .
\eeq
Following Darboux \cite{Darb}, one can define suitably normalized tangent vectors
\beq \label{Xi}
\bX_i = \frac{D_i\bx}{T_i\gamma_i} 
\eeq
which possess the property that their variation in the second direction $j\not=i$
is proportional to $\bX_j$ only:
\beq \label{eq:dLap3}
D_j\bX_i = (T_j\beta_{ji})\bX_j \; \; , \; \; i\not=j \;
\eeq
where 
\beq \beta_{ji} = \frac{D_i\gamma_j}{T_i\gamma_i} \; \; \;\; , \; \; \; 
 i\not=j \; \;  .
\eeq
The compatibility condition written in terms of the new fields $\beta_{ij}$
takes the following form
\beq \label{eq:dDarb3}
D_j\beta_{ik} = \beta_{ij}(T_j\beta_{jk}) \; \; , \; \; 
\; \; \; i\not=j\not= k\not= i \; .
\eeq
As it was mentioned in the introduction, the matrix version of equations 
(\ref{eq:dpDarb2})--(\ref{eq:dDarb3}), for $N=3$, has been derived in \cite{BoKo},
through a $\bar{\partial}$ dressing approach, without a geometric interpretation,
as an integrable discrete generalization of the DZM equations. 

\section{Geometric Construction of the Lattice}
We conclude the Letter with some remarks about the constructability of the 
quadrilateral lattices. We discuss also the meaning of some objects introduced
in the last Section.

Given two discrete curves $\bx_1(n_1)$ and $\bx_2(n_2)$ in $\RR^M$ meeting in 
a point $\bx_1(0)=\bx_2(0)$, and given two functions $A_{12}$ and $A_{21}$ 
defined on $\ZZ^2$, these data allow to construct uniquely the discrete 
quadrilateral surface. Equivalently, one can use instead of 
$A_{12}$ and $A_{21}$ the functions $\gamma_{1}$ and $\gamma_{2}$. 
Once the functions $\gamma_i$, $i=1,2$ are given, the 
vectors $\bX_{i}$ are also defined in the
points of the discrete quadrilateral surface according to formula (\ref{Xi}).

The construction of the vectors $\bX_{i}$ for a given 2-dimensional 
quadrilateral lattice (2DQL)
is shown in the Fig. 3. For example, $D_2\bX_1$ (the change of $\bX_{1}$ in 
the second
direction) is proportional to $\bX_{2}$ only (the dashed line connecting
the arrows of $\bX_{1}$ and $T_2\bX_{1}$ is parallel to $D_2\bX$).

\epsffile{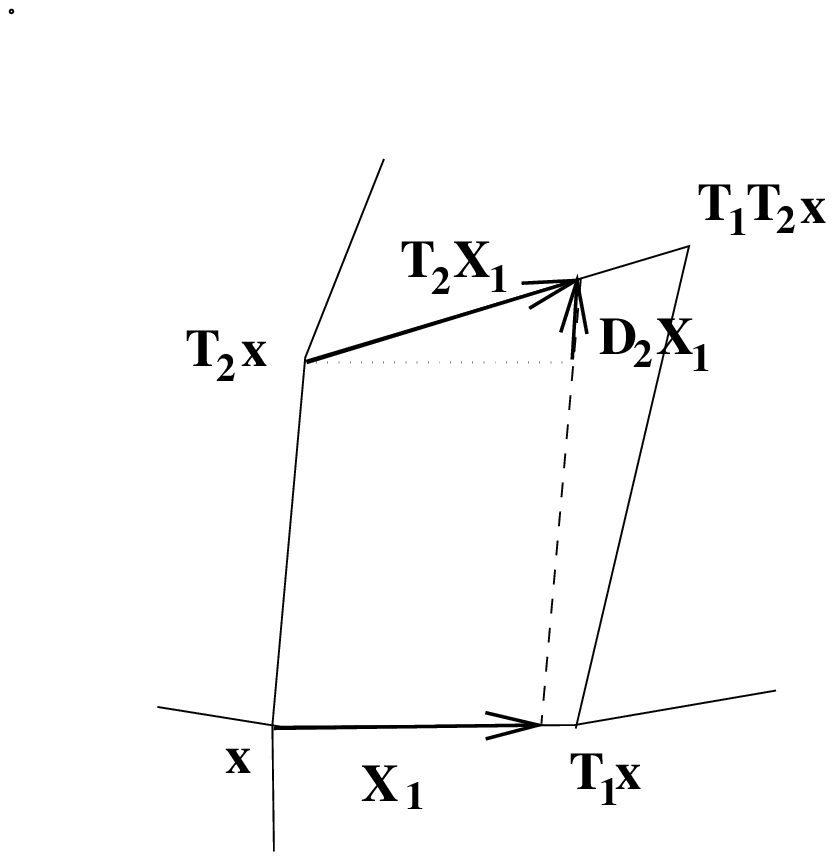}

Figure 3.

\bigskip

Let us consider now three discrete curves 
$\bx_1(n_1)$, $\bx_2(n_2)$ and $\bx_3(n_3)$ in $\RR^M$ ($M\geq 3$) meeting in 
a point $\bx_1(0)=\bx_2(0)=\bx_3(0)$. For any pair of curves: $\bx_i(n_i),\;
\bx_j(n_j)$, assigning two arbitrary functions $A_{ij}^{(0)}:\ZZ^2 \ra \RR$
we construct a $2$-dimensional (initial) QL.
It turns out that, starting from
these three initial discrete surfaces, one constructs uniquely the three 
dimensional quadrilateral lattice. The construction is based on the simple fact
that three different, non parallel, planes in a 3-dimensional space intersect,
in general,  in a single point. It may happen that the intersection point is 
at infninity, but we still have non ambiguous construction looking at the 
lattice in the projective space $\PP^M = \RR^M \cup \PP^{M-1}$.

In particular, the point $T_1T_2T_3\bx$ is the point of intersection of 
the three planes
$\langle T_1\bx , T_1T_2\bx, T_1T_3\bx \rangle$, 
$\langle T_2\bx , T_1T_2\bx, T_2T_3\bx \rangle$ and
$\langle T_3\bx , T_1T_3\bx, T_2T_3\bx \rangle$ in the three dimensional
subspace $\langle \bx, T_1\bx , T_2\bx, T_3\bx \rangle $ of $\RR^M$ 
(see Fig. 4).

\epsffile{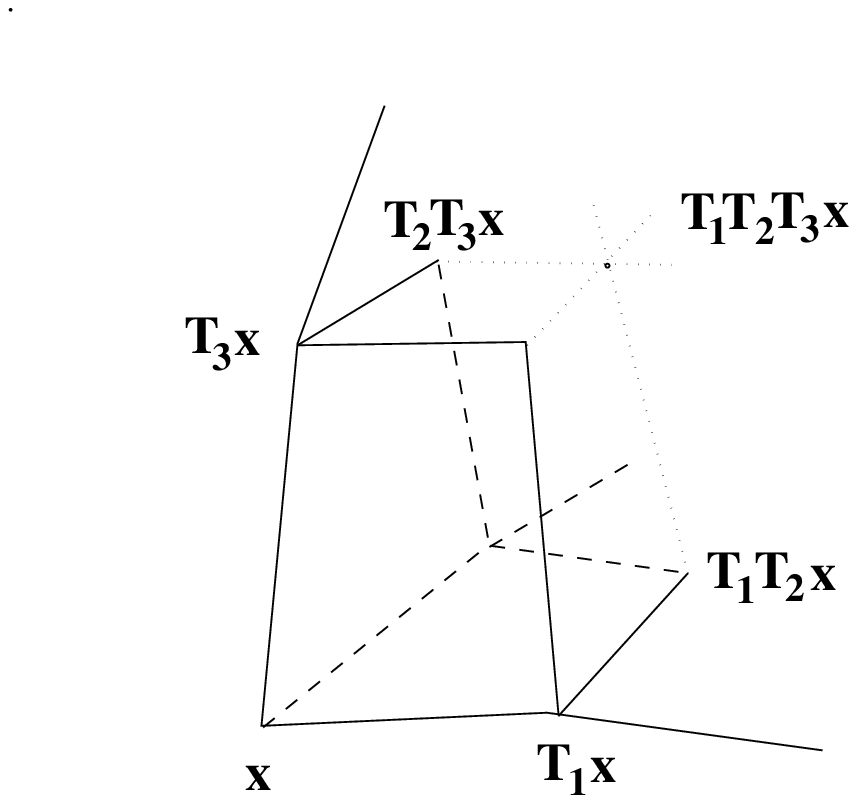}
 
Figure 4.

\bigskip

\nin The above construction involves only linear operations and can be
prolonged to build, out of the three initial quadrilateral surfaces, the whole
3-dimensional lattice.

Fixing the vectors $\bX_i$ (or the functions $\gamma_i$) on the initial curves 
allows to find their values in the points of the initial discrete quadrilateral 
surfaces, and then in the points of the whole lattice. Because of the planarity 
condition
there is no contradiction in the construction of the vectors $\bX_i$. 
For example,
the vector $T_2T_3\bX_1$ is obtained by the intersection of the line
$\langle T_2T_3\bx, T_1T_2T_3\bx \rangle$ with the plane parallel to 
$\langle \bx,T_2\bx, T_3\bx \rangle$ passing through the arrow of $\bX_1$ 
(see Fig. 5).

\epsffile{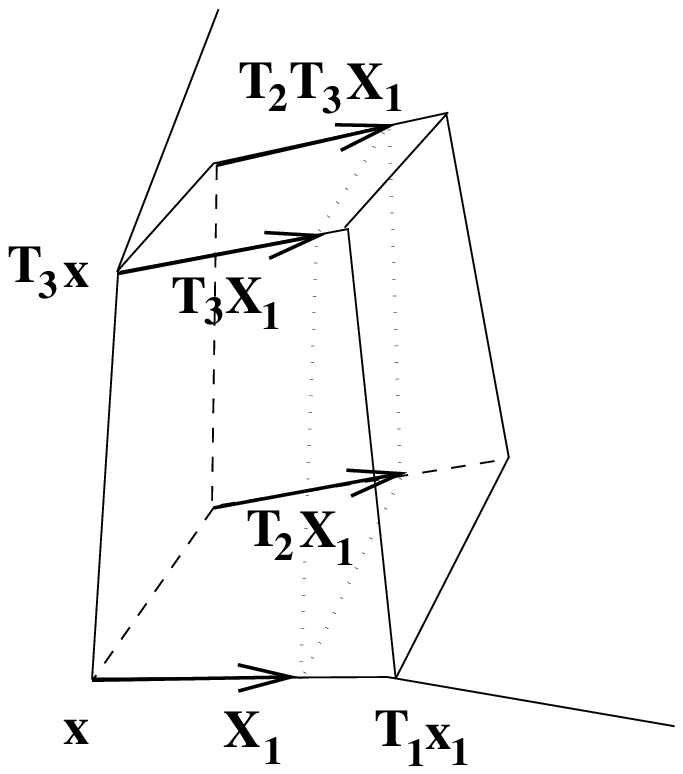}

Figure 5.

\bigskip

\nin To summarise, in order to construct uniquely the $3$-dimensional 
quadrilateral lattice, one has to give three arbitrary intersecting
quadrilateral surfaces or, equivalently, three arbitrary intersecting curves plus 
six arbitrary functions of two discrete variables
\beq
A_{ij}^{(0)}(n_i,n_j) \; \; , \; A_{ji}^{(0)}(n_i,n_j) \; \; \; , \; \; 
1\leq i < j \leq 3 \; \; .
\eeq
The construction outlined above allows then to build uniquely the whole
3-dimensional lattice or, equivalently, the functions 
\beq A_{ij}(n_1,n_2,n_3) \; \; , \; \; i\not= j\; \; , \; \; i,j=1,2,3
\eeq
which solve the MQL equation and satisfy the following boundary conditions:
\beq A_{12}(n_1,n_2,0) = A_{12}^{(0)}(n_1,n_2) \; \; , \; \; \dots \; ,
A_{32}(0,n_2,n_3)=A_{32}^{(0)}(n_2,n_3) \; \; .
\eeq
If, instead of the $A_{ij}$'s, we use the data $\gamma_i$'s,
then our arbitrary initial functions are  $\gamma_1^{(0)}(n_1,n_2)$ , 
$ \gamma_1^{(0)}(n_1,n_3)$,  
$\gamma_2^{(0)}(n_1,n_2)$, $\gamma_2^{(0)}(n_2,n_3)$,  
$\gamma_3^{(0)}(n_1,n_3)$ and $\gamma_3^{(0)}(n_2,n_3)$.

\bigskip

Let us apply the same procedure to the case $N=4$. Given 
four discrete curves and six, constructed from them, arbitrary initial 2DQL's,
they give rise to the four unique 3DQL's
$\bx(n_1,n_2,n_3,0)$, $\bx(n_1,n_2,0,n_4)$, $\bx(n_1,0,n_3,n_4)$ and 
$\bx(0,n_2,n_3,n_4)$ in the way described above. Then 
the point $\bx(1,1,1,1)$ can be constructed, for example, out of the surfaces
$\bx(1,n_2,n_3,0)$, $\bx(1,n_2,0,n_4)$ and $\bx(1,0,n_3,n_4)$, i.e. it is the
intersection point of the three planes $T_1T_2P_{34}$, $T_1T_3P_{24}$ and 
$T_1T_4P_{23}$ in the three dimensional subspace 
$T_1P_{234}$ of $\RR^M$. Here $P_{ij}$ $(i\not=j)$ denotes the plane passing 
through the points $\bx$, $T_i\bx$ and $T_j\bx$, i.e.   
$P_{ij}=P_{ij}(\bx)=\langle \bx , T_i\bx, T_j\bx \rangle$, e. g., 
$T_1T_2P_{34}=\langle T_1T_2\bx , T_1T_2T_3\bx, T_1T_2T_4\bx \rangle$; 
analogously 
$P_{ijk}=\langle \bx, T_i\bx , T_j\bx, T_k\bx \rangle $ (all the indices are
distinct). 
The same point could be constructed, however, also in three 
other ways. In the following we show that the results coincide. 

First, we note that the plane $T_1T_3P_{24}$ is the intersection of the two 
three dimensional subspaces  $T_1P_{234}$ and $T_3P_{124}$ in  the
four dimensional space 
$P_{1234}=\langle \bx,$ $ T_1\bx, $ $T_2\bx ,$ $ T_3\bx, $ $T_4\bx \rangle$. 
Similarly $T_1T_2P_{34}=T_1P_{234}\cap T_2P_{134}$ and 
$T_1T_4P_{23}=T_1P_{234}\cap T_4P_{123}$.

\epsffile{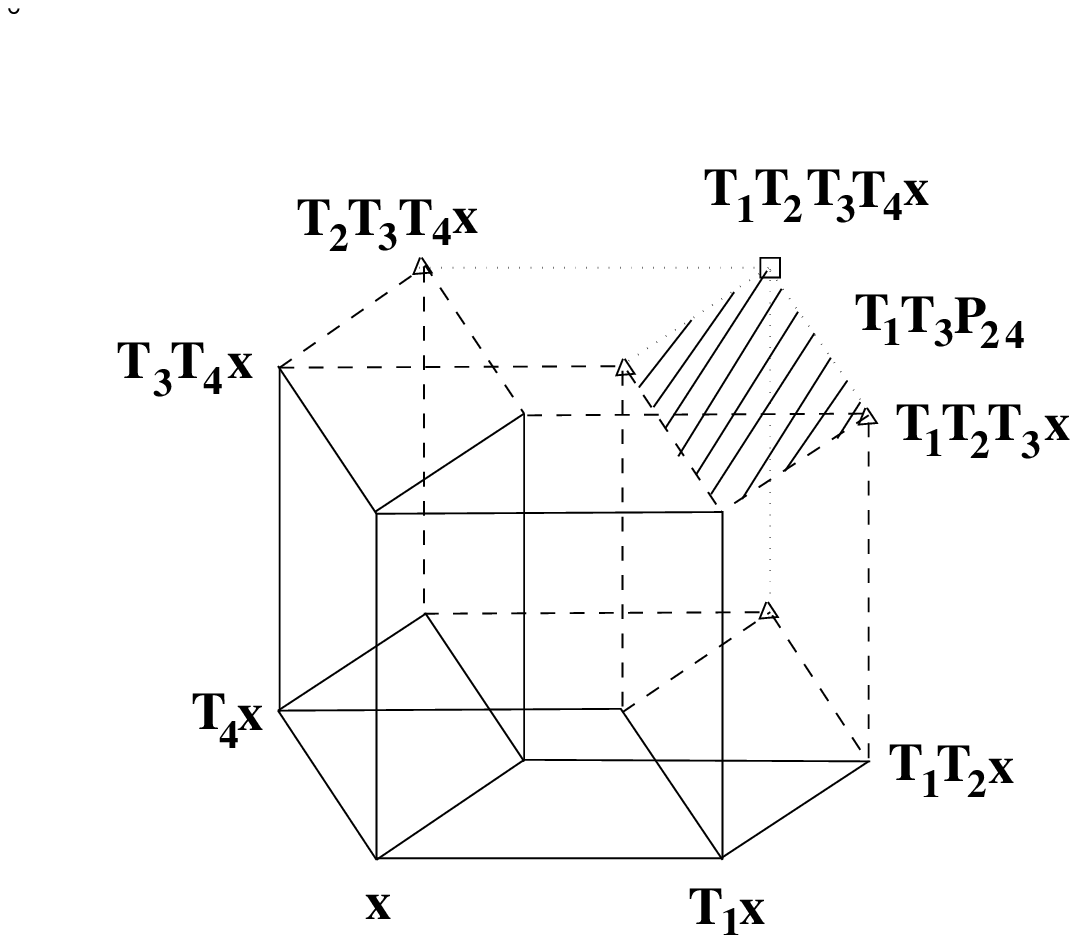}

Figure 6.

\bigskip

The point $T_1T_2T_3T_4\bx$ is then the unique intersection 
point of the four three dimensional subspaces $T_iP_{1.\check{i}.4}$, 
$i=1,\dots , 4$
(the symbol $1.\check{i}.4$ denotes the sequence of the natural numbers from 
$1$ to $4$ with the $i$th element removed)
of the four dimensional space $P_{1234}$; consequently, it is the 
intersection point of the four triplets $T_kT_iP_{1.\check{i}.\check{k}. 4}$, 
$k=1,\dots, 4$ (where the index $i$, $i=1,\dots , 4$, $i\not= k$, indicates
each element of the $k$th triplet) of the
two dimensional spaces in the corresponding three dimensional spaces
$T_kP_{1.\check{k}. 4}$
\beq P_{1234}\ni T_1T_2T_3T_4 \bx = \bigcap_{i=1}^{4}T_iP_{1.\check{i}. 4}
= \bigcap_{i=1,i\not=k}^{4}T_kT_iP_{1.\check{i}.\check{k}. 4}  \; \; \;  .
\eeq

The same  argument can be used to prove the compatibility of the 
construction for an arbitrary dimension $N$ of the lattice.
In the natural notation inherited from the example $N=4$
\beq P_{1\dots N}\ni T_1\dots T_N \bx = \bigcap_{i=1}^{N}T_iP_{1.\check{i}. N}
= \bigcap_{i=1,i\not=k}^{N}T_kT_iP_{1.\check{i}.\check{k}. N}  \; \; \; , 
\; \; k=1,\dots ,N \; \; 
\eeq
which allows to apply the mathematical induction.

To summarise all the geometric considerations of this Section, we conclude it
stating the general initial boundary--value problem for MQL's.

\bigskip

\nin {\it In order to construct uniquely the $N$-dimensional quadrilateral 
lattice, one has to give $\frac{N(N-1)}{2}$ arbitrary intersecting
quadrilateral surfaces or, equivalently, $N$ arbitrary intersecting curves (in 
general position) in $\RR^M$ plus $N(N-1)$ arbitrary functions of two discrete 
variables
\beq
A_{ij}^{(0)}(n_i,n_j) \; \; , \; \;  A_{ji}^{(0)}(n_i,n_j) \; \; \; , \; \; 
1\leq i < j \leq N \; \; \; .
\eeq
The {\bf linear construction} outlined above allows then to build uniquely the whole
$N$-dimensional lattice or, equivalently, the functions 
\beq A_{ij}(n_1,n_2\dots,n_N) \; \; , \; \; i\not= j\; \; , \; \; i,j=1,\dots,N
\eeq
which solve the MQL equation (\ref{eq:dDarb1}) and satisfy the following 
boundary conditions:}
\bea 
A_{ij}(0,\dots,n_i,\dots,n_j,\dots,0) & = & A_{ij}^{(0)}(n_i,n_j) \; \; ,  \\
A_{ji}(0,\dots,n_i,\dots,n_j,\dots,0) & = & A_{ji}^{(0)}(n_i,n_j) \; \; \; ,
\; \; 1\leq i < j \leq N \; \; . \nonumber
\eea

Note that the number of the arbitrary functions of two variables entering
into the general solution of the MQL equation agrees with that of the 
continuous case \cite{Darb}.

\bigskip

\nin {\bf Remark} All the geometric considerations of this Section can be 
reformulated in terms of linear equations (for example, a 3-dimensional
subspace of a 5-dimensional space is equivalent to a system of (5-3) linear
equations for 5 unknowns) and the corresponding theorems about their solutions.

\section{Final Remarks} 
It is important to emphasize that the MQL is, {\it by construction}, an
{\it integrable lattice}. Indeed, the linear constraints (\ref{eq:dLap})
(the {\it planarity} constraints) provide a well defined way to construct the 
lattice; therefore they 
charaterize completely the MQL and its integrability properties.

We are convinced that the simple geometric meaning of the MQL 
equations presented in this paper will allow to find for them 
many applications in all the branches of physics in which studies of 
multidimensional lattices are of importance.

We would like to point out that the projective picture
used in this paper should not be considered as a mathematical curiosity.
It allows to treat the points of the MQL which belong to the hyperplane
at infinity on an equal footing with the "usual" points of the affine
space. The presence of the points at infinity corresponds to
 singularities in the solution of the
MQL equation (\ref{eq:dDarb1}). It may be interesting to develop
this point of view in connection with the singularity confinement
property of integrable discrete systems \cite{singconf}.

\bigskip

\nin {\Large {\bf Acknowledgements}}

\bigskip

\nin A. D. would like to thank A. Sym for discussions and pointing out
reference \cite{Sauer}. He was supported 
partially by the Polish Committee of Scientific Research (KBN) under the Grant 
Number 2P03 B 18509.

\end{document}